\documentclass[12pt]{article}
\usepackage{graphicx}
\usepackage{amssymb}

\newcommand{\be}{\begin{equation}}
\newcommand{\ee}{\end{equation}}
\newcommand{\ba}{\begin{array}}
\newcommand{\ea}{\end{array}}
\newcommand{\bea}{\begin{eqnarray}}
\newcommand{\eea}{\end{eqnarray}}
\newcommand{\la}{\langle}
\newcommand{\ra}{\rangle}

\newcommand{\trace}{\mathop{\mathrm{Tr}}\nolimits}
\newcommand{\RR}{\mathbb{R}}
\newcommand{\CC}{\mathbb{C}}
\newcommand{\calH}{{\cal H }}
\newcommand{\calE}{{\cal E }}
\newcommand{\logs}[1]{\mathrm{log}^2{#1}}
\newcommand{\qed}{\begin{flushright} $\Box$ \end{flushright}}

\newtheorem{lemma}{Lemma}

\title{Upper bounds on entangling rates of bipartite Hamiltonians}
\author{Sergey Bravyi\\
\it IBM Watson Research Center,
Yorktown Heights, NY 10598, USA.}

\begin{document}
\maketitle

\abstract{
We discuss upper bounds on the rate at which unitary evolution governed by a non-local Hamiltonian
can generate entanglement in a bipartite system.
Given a bipartite Hamiltonian $H$ coupling two finite dimensional particles $A$ and $B$, the entangling rate is shown to
be upper bounded by $c\, \log{(d)}\,  \|H\|$, where $d$ is the smallest dimension of the interacting particles,
$\|H\|$ is the operator norm of $H$,
and $c$ is a constant close to $1$. Under certain restrictions on the initial state
we prove analogous upper bound for the ancilla-assisted entangling rate with a
constant $c$ that does not depend upon dimensions of local ancillas.
The restriction is that the initial state has at most two distinct Schmidt
coefficients (each coefficient may have arbitrarily large multiplicity).
Our proof is based on analysis of a mixing rate --- a functional measuring how fast
entropy can be produced if one mixes a time-independent state with a state evolving
unitarily.}

\section{Introduction}

Consider two remote parties Alice and Bob controlling finite-dimensional quantum systems $A$ and $B$.
Suppose $A$ and $B$
interact with each other according to a time-independent Hamiltonian $H$.
Unless $H$ is  a sum of local Hamiltonians, a unitary evolution $e^{iHt}$
is a non-local operation capable of
creating entanglement between $A$ and $B$.
The goal of the present paper is to get an upper bound on the rate at which
the entanglement between Alice and Bob can increase or decrease as a function of time.

\begin{figure}[t]
\label{fig:fig1}
\centerline{ \mbox{\includegraphics[height=4cm]{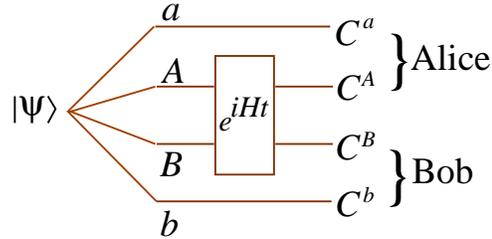}}}
\caption{Ancilla-assisted entangling.}
\end{figure}

We consider ancilla-assisted entangling, see Figure~\ref{fig:fig1}.
It means that Alice's laboratory consists of two subsystems: $A$ and $a$,
such that the Hamiltonian $H$ acts only on the subsystem $A$. Similarly, Bob's laboratory is
partitioned into $B$ and $b$, where $H$ acts only on the subsystem $B$.
The subsystems $a$ and $b$ are the local ancillas held by Alice and Bob.
In the ancilla-assisted entangling Alice and Bob start from a pure state $|\Psi\ra$ of
the composite system $aABb$ which may be already entangled.

Time evolution of the composite system $aABb$ is described by a unitary operator
$U(t)=I_a\otimes e^{iH_{AB} t} \otimes I_b$.
Thus the joint state of Alice and Bob always remains pure. Accordingly,
the entanglement between Alice and Bob at any time $t$ can be quantified by
entanglement entropy
\be
S(aA)=-\trace{ \rho_{aA}(t)\, \log{\rho_{aA}(t)}},
\quad
\rho_{aA}(t)=\trace_{Bb}  U(t)\, |\Psi\ra\la \Psi|\, U(t)^\dag.
\ee
The quantity we are interested in is {\it entangling rate}
\be\label{Gamma(Psi,H)}
\Gamma(\Psi,H)=\left. \frac{dS(aA)}{dt}\right|_{t=0}.
\ee
Understanding properties of the
 entangling rate is crucial for
optimal generation of entanglement~\cite{DVCLP00},
computing capacities of bidirectional quantum communication channels~\cite{BHLS02},
and for describing dynamics
of entanglement in quantum spin lattice models~\cite{BHV06}.

Our goal is to get an upper bound on $\Gamma(\Psi,H)$ that would not explicitly
depend upon dimensions of the local ancillas $a$ and $b$.
Upper bounds of this kind can be easily generalized
via the Tr\"otter
decomposition
to arbitrary multipartite
Hamiltonians decomposable into a sum of few-party interactions.

\subsection{Previous work}
The maximal entangling rate $\Gamma(H)=\max_\Psi \Gamma(\Psi,H)$ has been studied by many authors
for the case when $A$ and $B$ are qubits.
The optimal state $\Psi$ maximizing $\Gamma(\Psi,H)$ has been found by D\"ur, Vidal et~al~\cite{DVCLP00}
for a general Hamiltonian $H$ assuming $a=b=1$.
These authors have also observed that for some Hamiltonians
local ancillas are capable of increasing the maximal entangling rate
$\Gamma(H)$.
A powerful technique of getting upper bounds on $\Gamma(H)$
for arbitrarily large $a$ and $b$ was proposed
by Childs, Leung et~al~\cite{CLVV02}. It was used to identify a subclass of two-qubit Hamiltonians,
including the Ising interaction, for which the maximum  $\Gamma(H)$
does not depend upon $a$ and $b$ (and thus can be achieved without local ancillas).
An upper bound on the entangling rate of the Ising interaction has been also obtained by
Cirac, D\"ur et~al.~\cite{CDKL01} by showing that
the time evolution with the Ising Hamiltonian for time $t$ can be implemented by
LOCC protocol consuming $O(|t|)$ e-bits of pre-shared entanglement.

Some progress has been also achieved for larger dimensions of $A$ and $B$.
The technique of~\cite{CLVV02} has been
generalized by Wang and Sanders~\cite{WS02}
 to prove that
$\Gamma(H)\le \beta \approx 1.9123$ for any product Hamiltonian $H=H_A\otimes H_B$
if $H_A,H_B$ have eigenvalues $\pm 1$. This result applies to
arbitrary dimensions of $a$ and $b$.
An upper bound $\Gamma(H)\le \beta\, \| H\|$ for the ancilla-assisted
case has been proved for
arbitrary bipartite product Hamiltonians by Childs, Leung, and Vidal~\cite{CLV03}.
Finally,  it was shown by Bennett, Harrow et~al.~\cite{BHLS02} that
$\Gamma(H)\le c\, d^4\, \|H\|$, where $d=\min{(A,B)}$ and $c=O(1)$ does not
depend upon $a$ and $b$.
This result relies on a decomposition of an arbitrary bipartite Hamiltonian into a
sum of product Hamiltonians.
Thus for fixed $d$ and $\|H\|$ the entangling rate has a constant upper bound
independent on how large are dimensions $a$ and $b$.
 The authors of~\cite{BHLS02} also proved that
the supremum of $\Gamma(H)$ over
all dimensions of local ancillas $a,b$ coincides with the asymptotic capacity of $H$ to
generate entanglement by any protocol in which unitary evolution with $H$ is interspersed
with LOCC. It is unknown whether the supremum over $a$ and $b$ can be actually achieved for
finite dimensional ancillas.

It is not known whether $\Gamma(H)=\Gamma(-H)$, i.e., whether the maximal
entangling and disentangling rates of a given Hamiltonian $H$ coincide. The results by
Linden, Smolin, and Winter~\cite{LSW05} applicable to finite unitary operators suggest
that this equality might be wrong.

\subsection{Summary of results}
We start from observing that the unitary evolution with any Hamiltonian
$H_{AB}$ can not increase (decrease) the entanglement $S(aA)$ by more than $2\, \log{(d)}$,
where $d=\min{(A,B)}$, see~\cite{BHLS02}. This property can be called {\it small
total entangling}, as it says that the total increase (decrease) of entanglement
throughout the unitary evolution remains bounded as the dimensions of local ancillas $a$ and $b$
go to infinity. Is there an analogue of the small total entangling property for infinitely
small time intervals? The following conjecture was initially proposed by
Kitaev~\cite{pc}. We call it small incremental entangling.

\noindent
{\bf Small Incremental Entangling (SIE):} {\it There exists a constant $c=O(1)$ such that
\be
\label{SIE}
\Gamma(\Psi,H)\le c\, \|H\|\, \log{(d)}, \quad d\equiv \min{(A,B)}
\ee
for all dimensions $a$, $b$ and for all states $|\Psi\ra$ of a composite system $aABb$.}
\vspace{5mm}

It is not known whether SIE is true or false.
 Our first results is a proof of SIE for
the special case when $|\Psi\ra$ has at most two distinct Schmidt coefficients
with respect to a partition $aA\otimes Bb$ (each Schmidt coefficient may have
arbitrarily large multiplicity).
Our proof yields the constant $c=24$ although the actual value of $c$ might be much smaller.
The constraint on $|\Psi\ra$ serves technical purposes and was introduced in order to
make the problem tractable. It should be mentioned that the upper bound
Eq.~(\ref{SIE}) can be confirmed by
straightforward calculation if $a=b=1$ (no local ancillas), see Section~\ref{sec:no-ancillas}.
In this case the optimal pair $(\Psi,H)$
can be found explicitly. It turns out that the optimal state $|\Psi\ra$
has only two distinct Schmidt coefficients and thus falls into the category
that we consider, see Section~\ref{sec:no-ancillas} for details.
We don't know whether the optimal state has only two Schmidt coefficients in the
ancilla-assisted case.

It is known that in some cases local ancillas and pre-shared entanglement
can lead to counter-intuitive effects, such as
entanglement embezzling~\cite{DH03} or
 locking of classical correlations~\cite{DHLST03}.
In particular,
it was demonstrated by DiVincenzo, Horodecki et al.~\cite{DHLST03}
that sending a single qubit from Alice to Bob can increase their classical
mutual information by an arbitrarily large amount in the presence of local ancillas.
Thus SIE  may be violated if locking effects
also occur in the infinitesimal unitary transformations if one measures correlations by
entanglement entropy.
For future references let us give an explicit expression for the entangling rate that
can be easily obtained by computing the derivative in Eq.~(\ref{Gamma(Psi,H)}).
\be
\label{Gamma=}
\Gamma(\Psi,H)=-i\trace{
\left( I_a\otimes H_{AB}\, [\; \rho_{aAB},
\log{\rho_{aA}}\otimes I_B \; ]\right)}.
\ee

Our proof of SIE goes by getting an upper bound on a {\it mixing rate}. In order to define a mixing
rate, consider a probabilistic ensemble of (mixed) states $\calE=\{p_\alpha,
\rho_\alpha\}_{\alpha=0,1}$ defined on a finite-dimensional Hilbert space (lacking any tensor
product structure). Let $\rho=p_0\, \rho_0 + p_1\, \rho_1$ be the average state corresponding to
$\calE$. For any Hamiltonian $H$ define a time dependent state $\rho(t)=p_0\, \rho_0 + p_1\,
e^{iHt}\, \rho_1\, e^{-iHt}$. Define a mixing rate as
\be
\label{Lambda(E,H)}
\Lambda(\calE,H)
=\left. \frac{dS(\rho(t))}{dt}\right|_{t=0}.
\ee
Here $S(\rho(t))$ is the von~Neumann entropy of
$\rho(t)$. Basic properties of the von~Neumann entropy imply that $\bar{S}\le S(\rho(t)) \le
\bar{S}+h(p_0,p_1)$, where $\bar{S}= p_0\, S(\rho_0) + p_1\, S(\rho_1)$ is the average entropy for
the ensemble $\calE$ and $h(p_0,p_1)=-p_0\, \log{(p_0)} - p_1\, \log{(p_1)}$ is the Shannon
entropy. This property can be called {\it small total mixing} as it says that the total increase
(decrease) of the entropy throughout the unitary evolution goes to zero as
one of the probabilities $p_0$ or $p_1=1-p_0$ goes to zero.
Is there an analogue of the small total mixing property for
infinitely small time intervals? A naive generalization would be as follows.

\noindent
{\bf Small Incremental Mixing (SIM):} {\it There exists a constant $c'=O(1)$ such that
\be
\label{SIM}
\Lambda(\calE,H)\le c'\, \|H\|\, h(p_0,p_1),  \quad h(p_0,p_1)=-p_0\, \log{(p_0)} - p_1\, \log{(p_1)}
\ee
for any probabilistic ensemble $\calE=\{p_\alpha,\rho_\alpha\}_{\alpha=0,1}$.}
\vspace{5mm}

Here it is meant that the constant $c'$ is independent from the dimension of
the Hilbert space.
It is not known whether SIM is true or false.
Although SIM might seem completely unrelated to
ancilla-assisted entangling and
SIE, it turns out
that SIM is a stronger version of SIE. More strictly, we prove that
SIM with a constant $c'$ implies SIE
with a constant $c=4c'$, see Section~\ref{sec:reduction}.

Our second result is a proof of SIM
for the special case when $\rho=p_0\, \rho_0 + p_1\, \rho_1$ has at most two
distinct eigenvalues (each eigenvalue may have arbitrarily large multiplicity),
see Section~\ref{sec:SIM}.
We shall refer to eigenvalues obeying this constraint as a {\it binary spectrum}.
 Our proof yields $c'=6$, although the actual value of $c'$
might be smaller.

The connection between SIE to SIM described in Section~\ref{sec:reduction} has a peculiar
property that the number of distinct eigenvalues
of $\rho$ in SIM is exactly the same as the number of distinct Schmidt coefficients of the initial
state $|\Psi\ra$ in SIE. Thus a proof of SIM for the case when $\rho$ has a binary spectrum
implies SIE for the case when the initial state $|\Psi\ra$ has only two distinct Schmidt coefficients.
In the general case we prove that the mixing rate of any ensemble $\calE$ has an upper bound
$\Lambda(\calE,H)=O(\| H\|)$, so it does not explicitly depend on the dimension of the Hilbert space,
see Section~\ref{sec:SIM}.
For future references let us give an explicit expression for the mixing rate:
\be
\label{Lambda=}
\Lambda(\calE,H)
=-ip_1\trace{\left(
H\, [\; \rho_1,\log{\rho} \; ] \right)},
\quad
\rho=p_0\, \rho_0 + p_1\, \rho_1.
\ee

The paper is organized as follows. In Section~\ref{sec:no-ancillas} we find the
the maximal entangling rate and the optimal pair $(\Psi,H)$ for the case when
Alice and Bob do not use local ancillas. Section~\ref{sec:reduction} proves
that SIM implies SIE.
Section~\ref{sec:SIM} contains the main results of the paper. It proves SIM for
the case when $\rho$ has a binary spectrum. Section~\ref{sec:numerics} reports results
of numerical maximization aimed at verifying SIM. The data obtained in numerical
simulations are consistent with SIM.

\section{Maximal entangling rate in the absence of ancillas}
\label{sec:no-ancillas}
The maximal entangling rate can be easily found in the absence of local ancillas,
i.e., when  $a=b=1$. In this case we can always write the initial state of Alice and Bob
using the Schmidt decomposition $|\Psi\ra=\sum_{j=1}^d \sqrt{p_j}\, |j_A\ra\otimes |j_B\ra$
with  $d=\min{(A,B)}$.
Denote $\rho_A=\sum_{j=1}^d p_j \, |j\ra\la j|$ the reduced density matrix of Alice.
Using the general formula Eq.~(\ref{Gamma=}) for the entangling rate one gets
\be
\label{sec2:Gamma}
\Gamma(\Psi,H)=-\trace{\dot{\rho_A}\, \log{\rho_A}} =
-i\trace{\left( H\, [\, |\Psi\ra\la \Psi|,\log{\rho_A}\otimes I_B\, ]\right)}.
\ee
Since the entangling rate is a linear function of $H$ we can assume that $\|H\|=1$.
Using the fact that for any Hermitian operator $X$
\be
\label{sec2:1norm}
\max_{H\, : \|H\|=1} \trace{(H\, X)}=\trace{|X|}\equiv \|X\|_1,
\ee
one can carry out the maximization over $H$,
\be
\Gamma(\Psi):=\max_{H\, : \, \|H\|=1} \Gamma(\Psi,H) =
\|\; \;  [\,  \log{\rho_A} \otimes I_B, |\Psi\ra\la \Psi| \, ]\; \;  \|_1.
\ee
For any vectors $|\Psi\ra, |\Phi\ra$ one has the following identity
\be
\label{sec2:identity1}
\| \; |\Phi\ra\la \Psi| - |\Psi\ra\la \Phi| \; \|_1 = 2\sqrt{ \la \Psi|\Psi\ra \la \Phi|\Phi\ra - |\la \Psi|\Phi\ra|^2}.
\ee
Substituting $|\Phi\ra =(\log{\rho_A} \otimes I_B)\,|\Psi\ra$ one gets
\be
\label{sec2:F}
\Gamma(\Psi)= 2\sqrt{F(p)},
\quad F(p):=\sum_{j=1}^d p_j \, \logs{p_j} - \left(\sum_{j=1}^d p_j \, \log{p_j}\right)^2.
\ee
We can assume that all $p_j>0$ (otherwise replace $d$ by $d'<d$). Then the maximum
of $F(p)$ can be found by solving extremal point equations $\partial F/\partial p_j=0$.
After simple algebra one gets
\be
\label{sec2:extremum}
-\log{(2p_j)}=S\pm \sqrt{\gamma^2 +S^2}, \quad S=-\sum_{j=1}^d p_j\, \log{p_j},
\quad \gamma=\frac1{\ln{2}}.
\ee
(Here $\log$ and $\ln$ stand for base two and natural logarithm.)
The equality for the minus sign in the r.h.s. of Eq.~(\ref{sec2:extremum}) is possible
only for one value of $j$, since it implies $p_j\ge 1/2$. Let us agree that $p_1\ge 1/2$.
Then $-\log{(2p_j)}=S+ \sqrt{\gamma^2+S^2}$ for all $j=2,\ldots,d$, that is the state $\rho_A$
must have a binary spectrum
with multiplicities $1,d-1$.
 Introduce a variable $\lambda$ such that
\be
\label{sec2:spectrum}
p_1=\lambda, \quad p_2=\ldots=p_d=\frac{1-\lambda}{d-1}, \quad \frac12\le \lambda\le 1.
\ee
Let $|\Psi_\lambda\ra$ be the  state with Schmidt coefficients $p_1,\ldots,p_d$.
Using Eq.~(\ref{sec2:F}) the maximal entangling rate with the initial state $|\Psi_\lambda\ra$
can be written  as
\be
\label{sec2:Gamma(lambda)}
\Gamma(\Psi_\lambda)=2\sqrt{\lambda(1-\lambda)} \log{\left( \frac{\lambda(d-1)}{1-\lambda}\right)}.
\ee
The optimal value $\lambda=\lambda(d)$
and the optimal entangling rate $\Gamma_d=\Gamma(\Psi_{\lambda(d)})$
have to be found by maximizing Eq.~(\ref{sec2:Gamma(lambda)}) over $1/2\le \lambda\le 1$.
For example, if $d=2$, numerical maximization yields $\lambda(2)\approx 0.9168$ and
$\Gamma_2\approx 1.9123$. It coincides with the maximal entangling rate
of product two-qubit Hamiltonians found in~\cite{DVCLP00,CLVV02}.
It follows that under normalization condition $\|H\|=1$ product two-qubit
Hamiltonians are capable of generating entanglement with the largest rate.

One can easily infer from Eq.~(\ref{sec2:Gamma(lambda)}) that
$\lambda(d)\approx 1/2$ and $\Gamma_d\approx \log{d}$ for sufficiently large $d$.
It proves that $\Gamma(\Psi,H) =O(\|H\|\log{d})$ in the absence of local ancillas.
Moreover, in the limit of large $d$ one can explicitly write down the optimal state $|\Psi_{\lambda(d)}\ra$
and the optimal Hamiltonian $H_d$. Namely,
\be
\label{sec2:opt1}
|\Psi_{\lambda(d)}\ra\approx \frac1{\sqrt{2}}\, |1\ra\otimes |1\ra +
\frac1{\sqrt{2}}\, |\Phi^+\ra,
\quad |\Phi^+\ra =
\frac1{\sqrt{d-1}} \sum_{j=2}^d |j\ra\otimes |j\ra,
\ee
\be
\label{sec2:opt2}
H_d\approx -i(\, |\Phi^+\ra\la 1,1| - |1,1\ra\la \Phi^+|\, ).
\ee
Thus the optimal state is a superposition of a product state and a maximally entangled state
with locally orthogonal supports. The optimal Hamiltonian is a generator for a rotation in the
corresponding two-dimensional subspace.

\begin{figure}
\label{fig:fig2}
\centerline{
\includegraphics[height=5cm,clip]{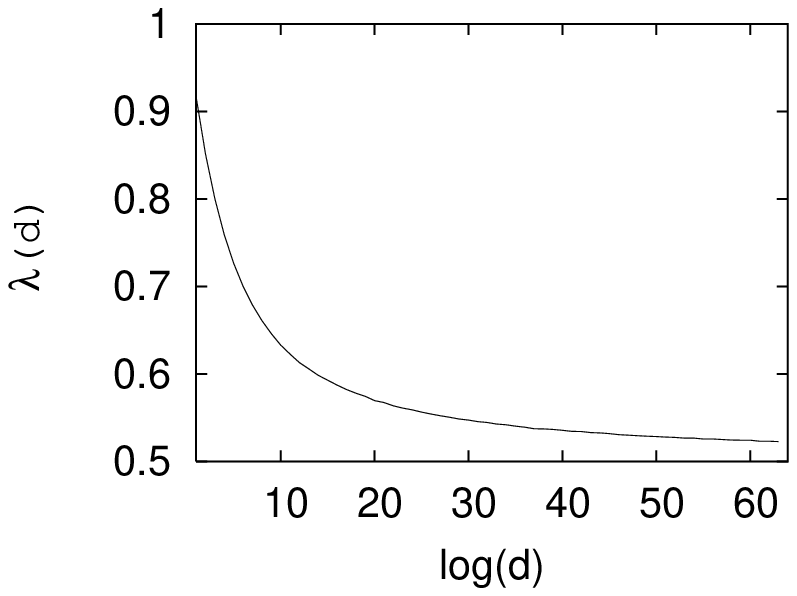}\hspace{1cm}
\includegraphics[height=5cm,clip]{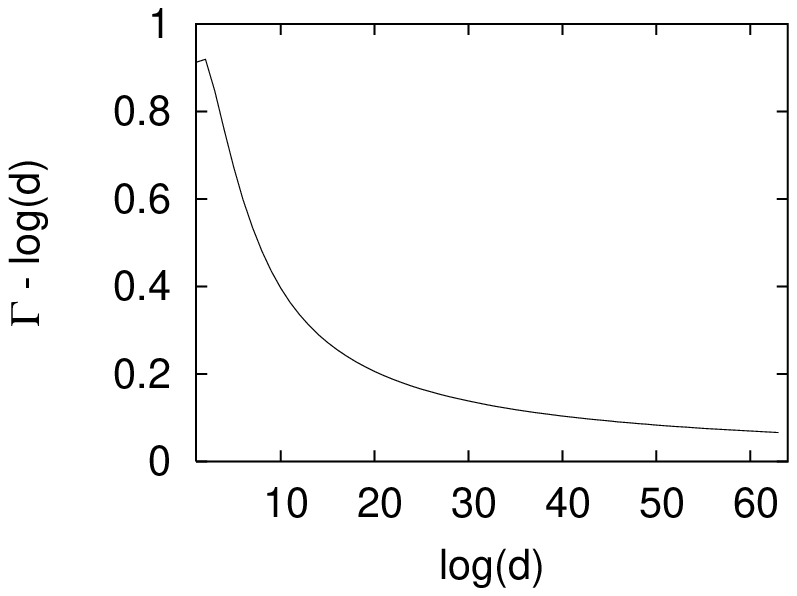}}
\caption{Optimal $\lambda$ (left) and entangling rate $\Gamma$ (right) as a function of $\log{d}$.}
\end{figure}

For a finite $d$ the optimal value $\lambda(d)$ and the corresponding entangling rate $\Gamma_d$
can be found numerically, see Figure~\ref{fig:fig2}. For large $d$ one has
$\lambda(d) \approx (1/2)(1+1/\ln{d})$, while entanglement entropy of $|\Psi_{\lambda(d)}\ra$
scales as $S_d \approx (1/2)\, \log{d}$.

\section{SIM implies SIE}
\label{sec:reduction}
In this section we assume that SIM is true and show that this assumption implies SIE.
Consider ancilla-assisted entangling
with dimensions $a,A,B,b$ and some initial state $|\Psi\ra$.
Let us assume that $B\le A$. Then SIE is equivalent to an upper bound
\be
\label{sec3:SIE}
\Gamma(\Psi,H)\le c\|H\| \log{B}.
\ee
Since the r.h.s. of this inequality does not depend on
$a$ and $A$, it is enough to prove Eq.~(\ref{sec3:SIE}) in the case $a=1$
since we can always extend $A$ to $A\otimes a$.
Now we have a tripartite system $ABb$ initially prepared in a pure state $|\Psi\ra$
and evolving under a Hamiltonian $H$ acting on $A$ and $B$.
Using formula Eq.~(\ref{Gamma=}) for the entangling rate one gets
\be
\label{sec3:Gamma}
\Gamma(\Psi,H)=
-i\trace{\left( H\, [\, \rho_{AB},\log{\rho_A}\otimes I_B\, ]\right)}=
-i\trace{\left( H\, [\, \rho_{AB},\log{\tau_{AB}}\, ]\right)},
\ee
where $\rho_A$ and $\rho_{AB}$ are the reduced states of $|\Psi\ra$, and
a state $\tau_{AB}$ is defined as
\be
\label{sec3:tau}
\tau_{AB}=\rho_A\otimes \frac{I_B}{B}.
\ee
In order to define a probabilistic ensemble $\calE$ needed for a reduction to SIM
we shall need the following lemma.
\begin{lemma}
\label{lemma:SIMtoSIE}
For any mixed state $\rho_{AB}$ there exists a mixed state $\mu_{AB}$ such that
\be
\label{sec3:decomposition}
\rho_A\otimes \frac{I_B}{B}=\frac1{B^2}\,  \rho_{AB} + (1-\frac1{B^2})\,  \mu_{AB}.
\ee
\end{lemma}
{\bf Proof:}
Since the partial trace is a linear operation, it suffices to prove the lemma for the
case when $\rho_{AB}=|\psi\ra\la \psi|$ is a pure state. The statement of the lemma is then equivalent to
inequality
\be
\label{sec3:lemma1}
|\psi\ra\la \psi| \le B\, \rho_A \otimes I_B.
\ee
Let $d$ be the Schmidt rank of $|\psi\ra$. Obviously, $d\le \min{(A,B)}$.
Consider a state $|I\ra=\sum_{j=1}^d |j_A\ra\otimes |j_B\ra$, where $\{ |j_A\ra\}$
and $\{|j_B\ra\}$ are the local bases of $A$ and $B$ that diagonalize $\rho_A$ and $\rho_B$.
Then
\[
|\psi\ra\la \psi|=(\rho_A^\frac12 \otimes I_B)\, |I\ra\la I|\, (\rho_A^\frac12 \otimes I_B).
\]
Taking into account that $\la I|I\ra=d$, one concludes that
$|I\ra\la I|\le d\, I_{AB}$.  Multiplying this inequality  by $\rho_A^\frac12 \otimes I_B$
on the left and on the right we get $|\psi\ra\la \psi|\le d\, \rho_A\otimes I_B$
which implies Eq.~(\ref{sec3:lemma1}).
\qed
Define an ensemble of states $\calE=\{p_\alpha,\tau_\alpha\}_{\alpha=0,1}$
such that
$\tau_0=\mu_{AB}$, $\tau_1=\rho_{AB}$, $p_0=1-B^{-2}$, $p_1=B^{-2}$. Here $\mu_{AB}$
is the state that appears in the decomposition Eq.~(\ref{sec3:decomposition}).
This ensemble has the average state $\tau=p_0\, \tau_0 + p_1\, \tau_1 =\rho_A\otimes I_B/B$.
Let $H$ be a Hamiltonian that appears  in Eq.~(\ref{sec3:Gamma}).
Assuming that SIM is true one gets
\be
\label{sec3:bound1}
\Lambda(\calE,H)\le c' \, h(B^{-2},1-B^{-2})\, \|H\| \le 4c' B^{-2}\, \log{B}\, \|H\|.
\ee
(Note that for any $x\le 1/2$ one has $h(x,1-x)\le 2x|\log{x}|$.)
On the other hand, formula Eq.~(\ref{Lambda=}) for the mixing rate
leads to
\be
\label{sec3:bound2}
\Lambda(\calE,H)=-ip_1\trace{\left(
H\, [\; \tau_1,\log{\tau}\; ]\right)} = -iB^{-2}\trace{\left(
H\, [\; \rho_{AB},\log{\tau_{AB}}\; ]\right)}.
\ee
Combining Eqs.~(\ref{sec3:bound1},\ref{sec3:bound2}) one arrives to
\[
-i\trace{\left(
H\, [\; \rho_{AB},\log{\tau_{AB}}\; ]\right)} \le 4c'\, \log{B}\, \|H\|.
\]
Comparing it with Eq.~(\ref{sec3:Gamma}) we infer that
$\Gamma(\Psi,H)\le 4c'\, \log{B}\, \|H\|$. We have proved SIE with
a constant $c=4c'$,
see Eq.~(\ref{sec3:SIE}).

The reduction Eq.~(\ref{sec3:tau}) has a peculiar property that the number of
distinct eigenvalues of $\tau_{AB}$ and $\rho_A$ are the same.
Since we have used identification $A=aA$, the number of distinct eigenvalues of $\rho_A$
coincides with the number of distinct Schmidt coefficients of $|\Psi\ra$.
Thus if we can prove SIM with $\rho(0)$ having
only two distinct eigenvalues, we will prove SIE
for the case when $|\Psi\ra$ has only two distinct Schmidt coefficients.

\section{Upper bounds on the mixing rate }
\label{sec:SIM}
Let us start from proving a weaker (compared to SIM) upper bound on the mixing rate.
\begin{lemma}
For any ensemble $\calE$ and Hamiltonian $H$ one has
\be
\Lambda(\calE,H)\le 2\|H\|.
\ee
\end{lemma}
{\bf Proof:}
Suppose $\calE=\{p_\alpha,\rho_\alpha\}_{\alpha=0,1}$, where the states
 $\rho_0, \rho_1$ and a Hamiltonian $H$ are defined on a $D$-dimensional Hilbert space.
Consider ancilla-assisted
entangling protocol shown on Figure~\ref{fig:fig1} with $a=1$, $A=D$, $B=2$, $b=D$.
Choose a Hamiltonian $\tilde{H}$ coupling $A$ and $B$ as
\[
\tilde{H}=H_A \otimes |1\ra\la 1|_B.
\]
Choose the initial state as
\[
|\Psi\ra = \sqrt{p_0}\, |0\ra_B \otimes |\phi_0\ra_{Ab} +\sqrt{p_1}\, |1\ra_B \otimes |\phi_1\ra_{Ab},
\]
where $|\phi_\alpha\ra \in \CC^D\otimes \CC^D$ is a purification of $\rho_\alpha$.
Then the state $|\Psi\ra$ evolves in time as
\[
|\Psi(t)\ra =\sqrt{p_0}\, |0\ra_B \otimes |\phi_0\ra_{Ab} +\sqrt{p_1}\, |1\ra_B \otimes
(e^{iH_A t}\otimes I_b)\,|\phi_1\ra_{Ab}.
\]
Accordingly, $\rho_A(t)=p_0\, \rho_0 + p_1\, e^{iHt} \rho_1 e^{-iHt}$.
Therefore
$\Lambda(\calE,H)=\Gamma(\Psi,\tilde{H})$.
Now we can use the upper bound on the entangling rate obtained in~\cite{WS02,CLV03}
for product Hamiltonians, namely $\Gamma(\Psi,\tilde{H})\le \beta\, \|\tilde{H}\|=\beta \, \|H\|$,
where $\beta\approx 1.9123\le 2$. The lemma is proved.
\qed
This lemma implies that the whole difficulty of proving SIM (if it is true) concerns the
limiting cases $p_0\to 0$ or $p_0\to 1$. Note also that $\Lambda(\calE,H)=\Lambda(\calE^T,-H)$,
where ensemble $\calE^T$ is obtained from $\calE$ by interchanging $p_0,\rho_0$ with $p_1,\rho_1$.
Thus we can assume that $0\le p_1\le 1/2$.

In the following we shall represent an ensemble $\calE=\{p_\alpha, \rho_\alpha\}_{\alpha=0,1}$
using the data $(p,\rho,\Pi)$, where
\be
\label{sec4:cv}
p=p_1, \quad \rho=(1-p)\, \rho_0 + p\, \rho_1, \quad \Pi=p\, \rho^{-\frac12}\, \rho_1\, \rho^{-\frac12}.
\ee
One can easily check that a triple $(p,\rho,\Pi)$ represents some
ensemble $\calE$ iff
\be
\label{sec4:admissible}
0\le p\le 1, \quad 0\le \Pi \le I, \quad \trace{\Pi \rho}=p.
\ee
Indeed, the only non-trivial statement is that $\Pi\le I$. It can be obtained from
inequality $p_1\, \rho_1\le \rho$ by multiplying it by $\rho^{-1/2}$ on the left and
on the right. Using the representation Eq.~(\ref{sec4:cv}) one can rewrite the mixing
rate as
\be
\label{sec4:mixing}
\Lambda(\calE,H)=-i\trace{\left( H\, \rho^{\frac12}\, [\; \Pi,\log{\rho}\; ]\,  \rho^{\frac12} \right)},
\ee
see Eq.~(\ref{Lambda=}). Given $\calE$, the optimal Hamiltonian $H$
maximizing the mixing rate can be
found using Eq.~(\ref{sec2:1norm}). Thus we have
\be
\label{sec4:Hout}
\Lambda(\calE):=\max_{H\, : \|H\|=1} \Lambda(\calE,H) =
\| \,\rho^{\frac12}\, [\; \Pi,\log{\rho}\; ]\,  \rho^{\frac12} \, \|_1.
\ee
It is worth mentioning that for given $\rho$ the upper bound Eq.~(\ref{SIM}) is true for sufficiently small
$p$. Indeed, applying
the triangle inequality for the trace norm to Eq.~(\ref{sec4:Hout}) one gets
$\Lambda(\calE)\le
2\|\, \rho^{\frac12} \Pi  \rho^{\frac12} \, \log{\rho} \, \|_1$.
Since $\|\, AB\, \|_1\le \|A\|\, \|B\|_1$ for any operators $A,B$ it follows that
$\Lambda(\calE)\le
2 \| \log{\rho}\| \|\, \rho^{\frac12} \Pi \rho^{\frac12} \, \|_1$. If $p$
is smaller than the smallest eigenvalue of $\rho$, one has $\|\log{\rho}\|\le |\log{p}|$
and thus
$\Lambda(\calE)
\le 2|\log{p}| \trace{(\Pi\rho)} =2p|\log{p}| \le 2h(p,1-p)$.

In the rest of the section we prove SIM under the assumption that $\rho$ has binary
spectrum,
\[
\rho=\sum_{j=1}^D \rho_j \, |j\ra\la j|, \quad \rho_j=\left\{
\ba{rcl} \lambda_1 &\mbox{if} & j=1,\ldots,m \\
\lambda_2 &\mbox{if} & j=m+1,\ldots,D, \\
\ea
\right., \quad \lambda_1\ge \lambda_2.
\]
Here $m$ can be arbitrary integer between $1$ and $D$.
Define projectors $R=\sum_{j=1}^m |j\ra\la j|$ and $R^\perp=I - R$ such that
\[
\rho =\lambda_1\, R + \lambda_2\, R^\perp, \quad
\rho^\frac12 = \sqrt{\lambda_1} \, R + \sqrt{\lambda_2}\, R^\perp,
\quad
\log{\rho} = \log{\lambda_1}\, R + \log{\lambda_2}\, R^\perp.
\]
After some algebra one gets the following identity
\be
\label{sec4:identity}
\rho^{\frac12}\, [\; \Pi,\log{\rho}\; ]\,  \rho^{\frac12} =
\log{\left(\frac{\lambda_1}{\lambda_2}\right)}
\,
\sqrt{\lambda_1\, \lambda_2}\,
[\; \Pi, R\; ].
\ee
In order to upper bound the trace norm of the commutator $[\Pi,R]$
we shall use the following.
\begin{lemma}[{\bf H\"older inequality}]
\label{lemma:holder}
Let $R$ be a projector and $\Pi$ be a
positive semi-definite operator. Then
\be
\|\, [\; \Pi,R \;] \, \|_1 \le 2\sqrt{ \trace{(\Pi\, R)}\, \trace{(\Pi\, R^\perp)}}.
\ee
\end{lemma}
\begin{lemma}
\label{lemma:canonical}
Let $R$ be a projector of rank $m$ and $\Pi$ be a Hermitian operator such that
$0\le \Pi \le I$. Then there exists a Hermitian operator $\Pi'$ such that\\
(i) $0\le \Pi' \le \Pi$,\\
(ii) $[\; \Pi,R\; ]=[\; \Pi',R \; ]$,\\
(iii) $\trace{\Pi'}\le m$.
\end{lemma}
We shall postpone the proof of the two lemmas above until the end of the section.
Let us apply Lemma~\ref{lemma:holder} to the commutator in Eqs.~(\ref{sec4:identity})
where $\Pi$ is replaced by $\Pi'$ from Lemma~\ref{lemma:canonical}.
Then we can upper bound the mixing rate in Eq.~(\ref{sec4:Hout}) as
\be
\label{sec4:Lambda<=}
\Lambda(\calE)\le
2\log{\left(\frac{\lambda_1}{\lambda_2}\right)}
\,
\sqrt{\lambda_1\, \lambda_2 \,
\trace{(\Pi'\, R)}\, \trace{(\Pi'\, R^\perp)}}.
\ee
In order to analyze the expression above
introduce new variables $m',p',x_1,x_2$ such that
\be
\label{sec4:variables}
m'=\trace{\Pi'},
\quad
p'=\trace{(\Pi'\rho)},
\quad
x_i=\lambda_i \, m'.
\ee
Expressing $\trace{(\Pi'\, R)}$ and $\trace{(\Pi'\, R^\perp)}$ in terms
of $m'$ and $p'$ one can rewrite Eq.~(\ref{sec4:Lambda<=}) as
\be
\label{sec4:Lambda<=2}
\Lambda(\calE)\le
2\log{\left(\frac{x_1}{x_2}\right)}
\,
\sqrt{ \frac{ x_1 \, x_2\, (p'-x_2)\, (x_1 - p')}{(x_1-x_2)^2}}\equiv g(x_1,x_2).
\ee
Let us find constraints on the variables $x_1,x_2$.
Noting that $\lambda_2\, I \le \rho \le \lambda_1\, I$ and taking the trace with $\Pi'$
one gets $x_2\le p' \le x_1$. Besides, condition~(iii) of Lemma~\ref{lemma:canonical}
implies that $m'\le m$ and thus $x_1\le m\lambda_1 \le \trace{\rho}=1$.
By obvious reasons one also has $x_2\ge 0$. Summarizing,
\be
\label{sec4:xconstraints}
0\le x_2 \le p'\le x_1 \le 1.
\ee
Now the problem of getting upper bound on the mixing rate reduces to
maximizing a function $g(x_1,x_2)$ in Eq.~(\ref{sec4:Lambda<=2})
under constraints Eq.~(\ref{sec4:xconstraints}).
We prove (see  Lemma~\ref{lemma:x1x2} at the end of
the section) that $\max_{x_1,x_2} g(x_1,x_2)\le 6\, p' |\log{p'}|$ as long as $p'\le 1/2$.
Using condition~(i) of Lemma~\ref{lemma:canonical} we get
$p'=\trace{(\Pi'\rho)}\le \trace{(\Pi\rho)}=p$. As was mentioned in the beginning of the
section, we can assume that $p\le 1/2$ and thus $p'\le p \le  1/2$.
Summarizing, we get
\[
\Lambda(\calE)\le 6\, p' |\log{p'}| \le 6\, p |\log{p}| \le 6 \, h(p,1-p),
\]
where we used the fact that a function $x|\log{x}|$ is monotone increasing on the
interval $[0,1/2]$. Thus we have proved SIM under the assumption that $\rho$
has a binary spectrum.

{\bf Proof of Lemma~\ref{lemma:holder}:}
H\"older inequality asserts that
\be
\label{sec4:holder}
\|\, XY\, \|_1 \le \sqrt{\trace{(X^\dag X)} \trace{(Y^\dag Y)}}
\ee
for any operators $X$ and $Y$, see~\cite{Bhatia}.
In order to choose proper $X$ and $Y$ let us use an identity
$[\, \Pi,R\, ]=R^\perp \Pi R - R \Pi R^\perp$ and triangle inequality for the
trace norm:
\[
\|\, [\; \Pi,R \;] \, \|_1 \le \| \,R^\perp \Pi R \, \|_1 + \| \,R \Pi R^\perp \, \|_1 =
2\| R^\perp \Pi R\|_1 =2\| \, (R^\perp \Pi^\frac12) (\Pi^\frac12 R) \, \|_1.
\]
Substituting $X=R^\perp \Pi^\frac12$ and $Y=\Pi^\frac12 R$ into Eq.~(\ref{sec4:holder})
and noting that $R$, $R^\perp$ are projectors we get the inequality stated in the lemma.
\qed

{\bf Proof of Lemma~\ref{lemma:canonical}:}
Any Hermitian operator $\Pi$ satisfying $0\le \Pi \le I$ can be written
as a convex combination of projectors:
\[
\Pi=\sum_\alpha p_\alpha \, \Pi_\alpha, \quad \Pi_\alpha^\dag\, \Pi_\alpha = \Pi_\alpha.
\]
Suppose we can find the
operator promised in the lemma for every projector in the sum, that is we can
find $\Pi_\alpha'$ such that
(i) $0\le \Pi_\alpha'\le \Pi_\alpha$, (ii) $[\Pi_\alpha',R]=[\Pi_\alpha,R]$,
and (iii) $\trace{\Pi_\alpha'}\le m$.
Then we can choose the desired operator $\Pi'$ as
$\Pi'=\sum_\alpha p_\alpha \Pi_\alpha'$.
Thus it suffices  to prove the lemma for the case when $\Pi$ is a projector.
Let $D$ be the dimension of the Hilbert space. Consider a direct sum decomposition
\[
\CC^D=\calH\oplus \calH^\perp,
\]
where $\calH$ is the range of $R$, so that $\mbox{dim}(\calH)=m$. Then we can write
\be
\label{matrices}
R=\left(\ba{cc} I & 0 \\ 0 & 0 \\ \ea\right), \quad
\Pi=\left(
\ba{cc} A & C \\ C^\dag & B \\ \ea \right).
\ee
Here $0\le A \le I$ and $0 \le B \le I$ are Hermitian operators on $\calH$ and $\calH^\perp$
and $C$ is some operator $C\, : \, \calH^\perp\to \calH$.
The requirement that $\Pi$ is a projector implies
\be
\label{pc}
A\, (I-A)=C C^\dag, \quad
B\, (I-B)=C^\dag C, \quad
C\, (I-B)=AC.
\ee
Consider a decomposition $A=A'\oplus A''$, $B=B'\oplus B''$ where
$0<A'<I$ and $0<B'<I$, while $A''$ and $B''$ have only eigenvalues $0,1$.
It follows from Eq.~(\ref{pc}) that
\[
A''\, CC^\dag = CC^\dag \, A''=0, \quad B''\, C^\dag C = C^\dag C \, B''=0.
\]
Thus $C^\dag\, |\psi\ra=0$ for any $|\psi\ra$ from the range of $A''$ and
$C\, |\phi\ra=0$ for any $|\phi\ra$ from the range of $B''$.
Accordingly, the projector $\Pi$ has the following block structure
\be
\label{big}
\Pi=\left( \ba{cccc}
A'' & 0 & 0 & 0 \\
0 & A' & C' & 0 \\
0 & (C')^\dag & B' & 0 \\
0 & 0 & 0 & B'' \\
\ea
\right).
\ee
Let $\Pi'$ be the central block in $\Pi$. Clearly $\Pi'$ is a projector and
$\Pi'$ satisfies conditions (i) and (ii) of the lemma.
It remains to check that $\trace{\Pi'}\le m$.
Let $m'$ be the dimension of the block $A'$. By definition, $m'\le m$.
Since $\Pi'$ is a projector, the operators $A',B',C'$ obey the same constraint
as  Eq.~(\ref{pc}), that is
\be
\label{pc1}
A'\, (I-A')=C (C')^\dag, \quad
B'\, (I-B')=(C')^\dag C, \quad
C'\, (I-B')=AC'.
\ee
 Since $0<A'<I$ and $0<B'<I$ we conclude that
$C'(C')^\dag$ and $(C')^\dag C'$ are non-singular matrices and thus the dimensions of the
blocks $A'$ and $B'$ both equal to $m'$.
Also from Eq.~(\ref{pc1}) we infer that
$A'=C'\, (I-B')\, (C')^{-1}$, that is the spectrum of $A'$
coincides with the spectrum of $I-B'$ including multiplicities.
Accordingly, $\trace{\Pi'}=\trace{A'}+\trace{B'}=m'\le m$.
Therefore $\Pi'$ satisfies all three conditions of the lemma.
\qed

\begin{lemma}
\label{lemma:x1x2}
Let $0\le q\le 1/2$ be a real number.
Consider a function
\[
g(x_1,x_2)=
2\log{\left(\frac{x_1}{x_2}\right)}
\,
\sqrt{ \frac{ x_1 \, x_2\, (q-x_2)\, (x_1 - q)}{(x_1-x_2)^2}}.
\]
Suppose $0\le x_2\le q$ and $q\le x_1\le 1$. Then $g(x_1,x_2)\le 6 q|\log{q}|$.
\end{lemma}

\noindent
{\bf Proof:}
Let us consider three cases:\\
{\bf case~1:} $2q\le x_1\le 1$, $q/4 \le x_2\le q$.\\
Then $\log{(x_1/x_2)}\le \log{(4/q)} \le 3|\log{q}|$. Using an upper bound
$x_2(q-x_2)\le q^2/4$
 we arrive to
\[
g(x_1,x_2)\le
 3q \, |\log{q}|\, \sqrt{\frac{x_1(x_1-q)}{(x_1-x_2)^2}} \le
3q\, |\log{q}|\, \sqrt{\frac{x_1}{(x_1-q)}} \le
\]
\[
\le 3q\, |\log{q}|\,\sqrt{\frac{2q}{2q-q}} \le 3\sqrt{2}q\, |\log{q}|.
\]
The last inequality follows from monotonicity of a function
$x_1/(x_1-q)$ on the interval $(q,\infty)$.

\noindent
{\bf case~2:} $q\le x_1\le 2q$, $q/4\le x_2\le q$.\\
First note that $\log{(x_1/x_2)}\le \log{(8)}=3$.
Take into account that
\[
\frac{(q-x_2)(x_1-q)}{(x_1-x_2)^2} \le \frac12.
\]
Therefore
\[
g(x_1,x_2) \le 3\sqrt{2}\, \sqrt{x_1x_2} \le 3\sqrt{2}\, \sqrt{(2q)q} \le 6q \le 6q \, |\log{q}|.
\]

\noindent
{\bf case~3:} $q\le x_1\le 1$,  $0\le x_2\le q/4$.\\
Introduce new variable $y$ such that $x_2=qy$, that is $0\le y\le 1/4$.
Then
\[
g(x_1,x_2)\le 2q\sqrt{y(1-y)}\, |\log{qy}|\,  k(x_1),
\]
where
\[
k(x_1)=\sqrt{\frac{x_1(x_1-q)}{(x_1-q/4)^2}}.
\]
One can check that $k(x_1)\le 2/\sqrt{3}$ for all $q\le x_1\le 1$.
Using inequality
\[
 \sqrt{y(1-y)}\, |\log{qy}| \le \sqrt{y}\, |\log{qy}| =
\sqrt{y}\, |\log{q}| + 2\sqrt{y}\, |\log{\sqrt{y}}|
\]
and noticing that $t\, |\log{t}|$ is monotone increasing for $0\le t\le 1/2$ we conclude that
\[
\sqrt{y}\, |\log{qy}| \le \frac12 \, |\log{(q/4)}| \le  \frac{3}{2} \, |\log{q}|
\]
for any $0\le y\le 1/4$. Thus
\[
g(x_1,x_2)\le 2\sqrt{3} q\, |\log{q}|.
\]
Combining all three cases we get
\[
g(x_1,x_2)\le 6q\, |\log{q}|.
\]
\begin{flushright}
$\Box$
\end{flushright}

\section{Numerical maximization of the mixing rate}
\label{sec:numerics}
This section describes numerical simulations aimed at verifying SIM.
Let us start from  expression Eq.~(\ref{sec4:mixing}) for the mixing rate and the constraints
Eq.~(\ref{sec4:admissible}). It is convenient to represent the Hamiltonian
as $H=2K-I$. Note that $\|H\|\le 1$ iff $0\le K \le I$.
Then $\Lambda(\calE,H)=F(K,\Pi)$ where
\be
\label{sec5:objective}
F(K,\Pi)=
-2i\trace{\left( K\, \rho^{\frac12}\, [\; \Pi,\log{\rho}\; ]\,  \rho^{\frac12} \right)}=
2i\trace{\left( \Pi\, \rho^{\frac12}\, [\; K,\log{\rho}\; ]\,  \rho^{\frac12} \right)}.
\ee
For a fixed average state $\rho$
SIM is equivalent to an upper bound
\be
\label{sec5:SIM}
\max_{K,\Pi} F(K,\Pi)\le c' \, h(p,1-p),
\ee
where the maximization is subject to
\be
\label{sec5:admissible}
0\le K \le I, \quad 0\le \Pi \le I, \quad \trace{(\Pi\rho)}=p.
\ee
We found the maximum of $F(K,\Pi)$ numerically for the average states
\be
\label{embez}
\rho=\rho^{(D)}=Z^{-1}\, \sum_{j=1}^D j^{-1} |j\ra\la j|, \quad Z=\sum_{j=1}^D j^{-1},
\ee
with the dimension $D=4,8,16,32$ and for several values of $p$ between $0$ and $1/2$,
see Figure~3 (as was mentioned in Section~\ref{sec:SIM}, it is enough
to consider $p$ between the smallest eigenvalue of $\rho$ and $1/2$).

The motivation for this particular choice of the average state comes from
the fact that the states $\rho^{(D)}$  can ``embezzle" any other mixed state
for sufficiently large $D$, see~\cite{DH03}.
More strictly, it was proved in~\cite{DH03} that for any state $\rho$
there exists an isometry $V$ such that
$\rho^{(D)} \approx V^\dag\, (\rho^{(D)}\otimes \rho)\, V$ in the sense that
fidelity between the two states goes to $1$ as $D$ goes to infinity.
In particular this is true for the optimal state $\rho$ corresponding to
the optimal ensemble $\calE=(p,\rho,\Pi)$ maximizing the mixing rate in Eq.~(\ref{sec4:Hout})
for a fixed $p$. On the other hand,
the state $\rho^{(D)}\otimes \rho$ is at least as good as $\rho$ as far as the mixing rate is concerned.
Indeed, one can easily verify that
the mixing rate
for the ensemble $(p,\rho,\Pi)$ is the same as the mixing rate for the ensemble
$(p,\tilde{\rho},\tilde{\Pi})$, where
$\tilde{\rho}=\rho^{(D)}\otimes \rho$ and $\tilde{\Pi}=I\otimes \Pi$.
 Thus a verification of SIM for
the family of states $\rho^{(D)}$ is a good test for general validity of SIM.

The data obtained in the numerical maximization of $F(K,\Pi)$
are presented in Figure~3. They are consistent with
the conjecture Eq.~(\ref{sec5:SIM}) with a constant $c'=1$.
The numerical algorithm that we used is based on reformulation of the maximization
problem as a semidefinite program. The details of the algorithm are
described in Appendix~A.
\begin{figure}[h]
\label{fig:plots}
\centerline{ \mbox{\includegraphics[height=12cm,angle=-90]{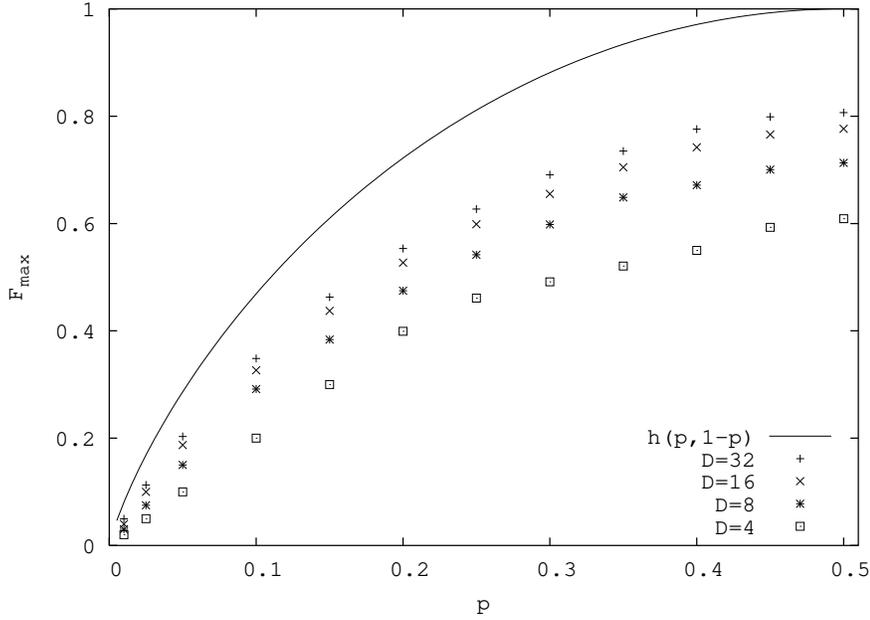}}}
\caption{Maximal mixing rate $F_{max}=\max F(K,\Pi)$ (vertical axis) computed for a fixed
average state $\rho=\rho^{(D)}$, see Eq.~(\ref{embez}), and fixed $p$ (horizontal axis).
The maximization was performed for $D=4,8,16,32$.
The upper curve shows the Shannon entropy $h(p,1-p)=-p\log{p}-(1-p)\log{(1-p)}$.}
\end{figure}

\vspace{1cm}

\noindent
{\bf Acknowledgements:} The author thanks Alexei Kitaev for helpful discussions.
 Numerous useful comments from Charles Bennett and
John Smolin are acknowledged.
This research was supported by the NSA and the ARDA through ARO contract number
W911NF-04-C-0098.

\section*{Appendix A}

Maximization of the objective function $F(K,\Pi)$, see Eq.~(\ref{sec5:objective}), over $K$ for fixed $\Pi$
or vice verse is a semi-definite program that can be solved very efficiently.
In order to find the global maximum $F_{max}$ of $F(K,\Pi)$ we used a sequence of $N\gg 1$ alternating
single-variable maximizations over $K$ and over $\Pi$.
To reduce the probability of being trapped in a local maximum, the whole
procedure was repeated $M\gg 1$ times with a random choice of initial pair $K,\Pi$.
The algorithm outputs the maximal found value of $F$ among all $M$ rounds.
The data presented on Figure~3
were obtained with a choice of parameters $N=32$ and $M=128$.
Based on fluctuations in the values of $F_{max}$ found in different rounds
we estimate the precision of the algorithm as $\delta \le 10^{-2}$.

Maximization of $F$ over $K$ for fixed $\Pi$ reduces to diagonalization of a Hermitian
traceless operator $Z:=-2i\rho^{\frac12}\, [\; \Pi,\log{\rho}\; ]\,  \rho^{\frac12}$, namely
\[
\max_{0\le K \le I} F(K,\Pi) = \max_{0\le K \le I} \trace{(Z\, K)}= \frac12\, \|Z\|_1.
\]
The optimal $K$ can be chosen as a projector onto the positive eigensubspace of $Z$.

Maximization of $F$ over $\Pi$ for fixed $K$ reduces to a semidefinite program. Define
a Hermitian traceless operator $X=2i\rho^{\frac12}\, [\; K,\log{\rho}\; ]\,  \rho^{\frac12}$.
Then we have to find
\be\label{primal}
F_{max}(p,\rho)=\max_{0\le \Pi\le I, \; \trace{(\Pi\,\rho)}=p} \trace{(X\, \Pi)}.
\ee
In practice it is more convenient to solve semidefinite program dual to Eq.~(\ref{primal})
as it reduces to diagonalization of operators and minimization of a convex function of
one real variable which can be done by the gradient methods.
The following lemma shows the connection between the primal and the dual
problems.
\begin{lemma}
\label{lemma:numerics}
Denote
\be
\label{lambda0}
\lambda_0=\min_{\lambda\in \RR}
 \left( \lambda (p-\frac12) + \frac12\|X-\lambda \, \rho\|_1 \right).
\ee
Let $P_+$ and $P_0$ be orthogonal projectors onto the
positive eigensubspace and zero eigensubspace
of an operator $X-\lambda_0\, \rho$ respectively.
Then $F_{max}(p,\rho)=\lambda_0$ and the optimal operator $\Pi$ can be  chosen as
\be
\Pi=P_+ + x\, P_0,
\ee
where the coefficient $x$ is determined from $\trace{(\Pi\, \rho)}=p$.
\end{lemma}
{\bf Proof:}
Let us find semidefinite program dual to Eq.~(\ref{primal}). The equality $\trace{(\rho\, \Pi)}=p$
gets Lagrangian multiplier $\lambda\in \RR$. The inequalities $\Pi\le I$ and $-\Pi\le 0$
get Lagrangian multipliers $A\ge 0$ and $B\ge 0$.
If one can choose $\lambda,A,B$ such that $X=\lambda\, \rho + A -B$ (the gradient of the objective
function is a convex combination of gradients of the constraints),
one gets an upper bound
\be
\label{sec5:gap}
\trace{(\Pi\, X)} = \lambda\, \trace{(\Pi\, \rho)} + \trace{(A\, \Pi)} - \trace{(B\, \Pi)}
\le \lambda\, p + \trace{A}.
\ee
The duality principle asserts that this upper bound is tight
if either the primal or the dual problems are strictly feasible (all inequalities
can be made strict). In our case the primal problem Eq.~(\ref{primal}) is strictly feasible for any $0<p<1$.
Indeed, take $\Pi=p\, I$. Then
$0<\Pi<I$ and $\trace{(\Pi\, \rho)}=p$.
Thus $F_{max}(p,\rho)$ coincides with the solution of the dual problem
\be\label{dual}
\mbox{minimize} \, (\lambda\, p + \trace{A}) \quad \mbox{subject to}
\quad
\left\{ \ba{c}
A\ge 0, \; B\ge 0, \\
X=\lambda\, \rho + A -B \\
\ea
\right.
\ee
Let us carry out the minimization in two stages: first minimize the objective function
over $A$ and then minimize the resulting (non-linear) function of $\lambda$.
Let $\lambda,A,B$ be any feasible solution of the dual problem. The
triangle inequality for the trace norm implies that
\be
\label{sec5:aux1}
\trace{A} + \trace{B} \ge \|X-\lambda\, \rho\|_1.
\ee
Since $X$ is traceless, we also have
\be
\label{sec5:aux2}
\trace{A} - \trace{B}=-\lambda.
\ee
Adding Eqs.~(\ref{sec5:aux1},\ref{sec5:aux2}) together we get
\be
\label{sec5:aux3}
\trace{A} \ge \frac12 \left(\|X-\lambda\, \rho\|_1  -\lambda \right).
\ee
The equality here is achieved when $A$ and $B$ are the positive
and the negative parts of $X-\lambda\, \rho$.
Thus we get
\be
\label{mina}
F_{max}(p,\rho)=\min_{\lambda\in \RR} \left( \lambda (p-\frac12) + \frac12\|X-\lambda\, \rho\|_1 \right).
\ee
The objective function is a convex one. It grows as $(1-p)|\lambda|$ for $\lambda\to -\infty$ and
as $p\, \lambda$ for $\lambda\to \infty$. Therefore the minimum is achieved at some finite $\lambda$
which we denote $\lambda_0$.
Equality in Eq.~(\ref{sec5:gap}) is possible only if
\[
\trace{(I-\Pi)\, A}=0, \quad \trace{(\Pi\, B)}=0, \quad \trace{(\Pi\rho)}=p.
\]
One can choose a solution as $\Pi=P_+ + x\, P_0$, where
$x$ is to be found from the constraint $\trace{(\Pi\rho)}=p$.
\qed

\end{document}